\documentclass[aps,epsfig]{revtex4}

\usepackage{epsfig}

\begin{document}


\title{\bf Viscosity in the escape-rate formalism}

\author{S. Viscardy and P. Gaspard\\
{\em Center for Nonlinear Phenomena and Complex Systems,}\\
{\em Universit\'e Libre de Bruxelles,}\\
{\em Campus Plaine, Code Postal 231, B-1050 Brussels, Belgium}\\}

\date{\today}

\begin{abstract}
We apply the escape-rate formalism to compute the shear viscosity in terms
of the chaotic properties of the underlying microscopic dynamics.  
A first passage problem is set up for the escape of the Helfand moment associated with
viscosity out of an interval delimited by absorbing boundaries.  At the microscopic level
of description, the absorbing boundaries generate a fractal repeller.  The fractal
dimensions of this repeller are directly related to the shear viscosity and the Lyapunov
exponent, which allows us to compute its values.  We apply this method to the
Bunimovich-Spohn minimal model of viscosity which is composed of two hard disks in elastic
collision on a torus.  These values are in excellent agreement with
the values obtained by other methods such as the Green-Kubo and
Einstein-Helfand formulas.  
\end{abstract}

\maketitle

\vskip 0.3 cm

\hspace*{16mm}{PACS numbers: 05.45.Ac; 05.45.Df; 05.45.Jn; 05.60.-k}]

\section{Introduction}

The link between the irreversible phenomena governed by macroscopic
equations such as
the Navier-Stokes equations and the microscopic reversible dynamics of the
atoms and
molecules is a fundamental problem.  In this context, it has been shown that
typical many-body systems of interacting particles present a chaotic dynamics
\cite{livi,posch-hoover,vanbei-dorfman,gasp-vanbei}.  This microscopic
chaos develops a
sensitivity to initial conditions over a time scale of the order of the
intercollisional time of
the atoms and molecules.  This sensitivity to initial conditions is
characterized by positive
Lyapunov exponents which are the rates of exponential separation between
some reference and
perturbed trajectories of the system.  The sensitivity to initial
conditions results into a
huge dynamical randomness characterized by a positive Kolmogorov-Sinai (KS)
entropy
per unit time given by the sum of positive Lyapunov exponents if the system
is at
equilibrium:
\begin{equation}
h_{\rm KS} = \sum_{\lambda_i>0} \lambda_i \; ,
\label{Pesin}
\end{equation}
an identity known as Pesin's equality \cite{eckmann-ruelle,ott}.
This dynamical chaos provides an efficient mechanism of randomization of the
different observable quantities such as the microscopic currents associated
with the
transport properties.

For a Hamiltonian-type microscopic dynamics, a connection can be
established between
dynamical chaos and the transport properties thanks to the escape-rate
formalism
\cite{gasp-nicolis,dorf-gasp,gasp-dorf,gasp-book,dorf-book}.  In this
formalism, the gap between
the kinetic time scale of chaotic properties and the hydrodynamic time
scale of transport
properties is bridged by linking the transport coefficients to {\it
differences} between chaotic
properties.  In this formalism, the following formula has been derived for
viscosity
\cite{dorf-gasp}
\begin{equation}
\eta = \lim_{\chi\to\infty} \left( \frac{\chi}{\pi}\right)^2 \left(
\sum_{\lambda_i>0} \lambda_i - h_{\rm KS} \right)_{{\cal F}_{\chi}} \; ,
\label{esc.eq}
\end{equation}
where the difference between the sum of positive Lyapunov exponents and the KS
entropy is nonvanishing because the system is here under {\it nonequilibrium}
conditions.  These nonequilibrium conditions select the trajectories of the
many-body
system which do not escape out of a phase-space region specific to the
transport
property of interest.  This phase-space region is defined by requiring that the
Helfand moment associated with the transport property remains bounded in an
interval
of extension $\chi$.  In the limit $\chi\to\infty$, the nonequilibrium
condition is
progressively relaxed and the sum of positive Lyapunov exponents as well as
the KS entropy approach their equilibrium value satisfying Pesin's equality
(\ref{Pesin}).  Under nonequilibrium conditions, Pesin's equality is not
satisfied
and the difference gives the rate of escape of trajectories out of the
aforementioned
phase-space region \cite{eckmann-ruelle,kantz-grassberger}.  This region
contains a fractal
repeller ${\cal F}_{\chi}$ composed of trajectories which escape neither in
future nor in past.
The escape rate is characteristic of this fractal repeller and is related
to the transport
coefficient, leading to the formula (\ref{esc.eq}).  The escape-rate
formalism has
already been applied to the transport property of diffusion
\cite{gasp-baras} as well as to
reaction-diffusion processes \cite{claus-gasp,claus-gasp-vanbei}.

The purpose of the present paper is to apply the escape-rate formalism to
viscosity.
The system we use as a vehicle of our study is a minimal model of viscosity
previously analyzed by Bunimovich and Spohn \cite{spohn-buni}.  The minimal
models of transport
are of special interest because they are the simplest possible models
already presenting a
positive and finite transport coefficient.  It is known that a minimal
model should
contain only one particle for diffusion, two particles for viscosity, and three
particles for heat conduction \cite{spohn-buni}.  For viscosity, we
therefore consider here the
model composed of two hard disks moving on a torus and undergoing elastic
collisions.  In a
previous paper \cite{premier-article}, we have described this model and
some of its properties for a hexagonal and a square geometries.  Our aim is
here to
compute viscosity thanks to the escape-rate formalism and to show the
equivalence
with the results of the Green-Kubo and Einstein-Helfand formulas already
obtained in
the previous paper \cite{premier-article}.

In the present paper, we compute the viscosity by using the chaotic and
fractal properties
of the repeller.  We use a variant of Eq. (\ref{esc.eq}) in which the
difference between positive
Lyapunov exponents and KS entropy is given in terms of the Lyapunov
exponents and the
partial fractal dimensions of the fractal repeller.  Indeed, the
fractality of the repeller is a corollary of its chaoticity so that its fractal
dimensions are related to its KS entropy.  Accordingly, the knowledge of
the partial
dimensions allow us to evaluate the KS entropy.

The paper is organized as follows. In Sec. \ref{escape}, we develop the
escape-rate formalism
for shear viscosity.  In Sec. \ref{repeller}, we present the fractal
repeller of viscosity in the
two-disk model, which we compare with the fractal repeller of diffusion in
the Lorentz gas.  In
this way, we show that a specific fractal repeller is associated with each
transport property.
Finally, the chaotic and fractal properties are described in Sec.
\ref{chaos} where we compute
the viscosity coefficient from the positive Lyapunov exponent and the
fractal dimension of the
repeller.  Conclusions are drawn in Sec. \ref{conclusions}.


\section{VISCOSITY IN THE ESCAPE-RATE FORMALISM}
\label{escape}

\subsection{Helfand moment for viscosity}

In the previous paper \cite{premier-article}, we have shown that the shear
viscosity coefficient
can be computed with the Einstein-Helfand formula \cite{helf}
\begin{equation}
\eta = \eta_{xy,xy}=\eta_{yx,yx} = \lim_{t\to\infty} \frac{1}{2t} \;
\langle [ \tilde
G_{yx}(t)-\tilde G_{yx}(0)]^2 \rangle \; ,
\label{Einstein}
\end{equation}
where the Helfand moment is defined by
\begin{equation}
\tilde G_{yx} \equiv \sqrt{\frac{\beta}{V}} \; G_{yx} =
\sqrt{\frac{\beta}{V}} \left[ \sum_{a=1}^N
x_a(t) \; p_{ay}(t) - \sum_{a=1}^N \sum_s \Delta x_a^{(s)} \; p_{ay}^{(s)}
\; \theta(t-t_s)
\right]
\label{Helfand}
\end{equation}
for a system of $N$ particles of position ${\bf r}_a=(x_a,y_a,...)$ and
momentum ${\bf
p}_a=(p_{ax},p_{ay},...)$ moving in a domain delimited by periodic boundary
conditions.
As explained in the previous paper \cite{premier-article}, the particles
must satisfy the minimum
image convention, which requires the presence of the extra terms in the
Helfand moment
(\ref{Helfand}) involving the jumps $\Delta x_a^{(s)}$ of the particles to
fulfil the periodic
boundary conditions.  $t_s$ are the times of the jumps.
$p_{ay}^{(s)}=p_{ay}(t_s)$ is the
momentum at the time $t_s$ of the jump.  In Eq. (\ref{Einstein}), the average
$\langle\cdot\rangle$ is taken in the equilibrium microcanonical ensemble
for which
\begin{equation}
\beta = \frac{N}{k_{\rm B}T(N-1)} \; .
\label{invtemp}
\end{equation}

\subsection{First passage problem for viscosity}

The central object of the escape-rate formalism is the fractal repeller
which is composed of the
phase-space trajectories for which the Helfand moment fluctuates forever
(in the future and the
past) within some interval:
\begin{equation}
-\frac{\chi}{2} \leq \tilde G_{yx} \leq +\frac{\chi}{2} \; .
\label{abc}
\end{equation}
These trajectories are exceptional because the Helfand moment escape out of
this
interval for almost all the trajectories.  Therefore, the repeller has a
vanishing probability
measure in the phase space albeit it is typically composed of a
nonenumerable set of
trajectories.  The repeller thus typically forms a fractal in the phase space
\cite{gasp-dorf,gasp-book}.

We set up a first passage problem of the Helfand moment by introducing
absorbing boundaries at
$\tilde G_{yx}=\pm\frac{\chi}{2}$.  These absorbing boundaries in the space
of variations of the
Helfand moment correspond to equivalent absorbing boundaries in the phase
space of the system.
In the phase space, the absorbing boundaries delimit a domain which
contains the fractal repeller.
We consider a statistical ensemble of initial conditions taken inside this
domain and we run their
trajectories.  When a trajectory reaches the absorbing boundaries it escape
out of the
domain and it thus removed out of the statistical ensemble.

Under the forward time evolution, the remaining trajectories belong
to the stable manifolds of the repeller.  Under the backward time
evolution, the remaining
trajectories belong to the unstable manifolds of the repeller.  Under both
the forward and
backward time evolution, the remaining trajectories belong to the repeller
itself which is the
intersection of its stable and unstable manifolds \cite{gasp-dorf}.  For a
typical chaotic
dynamics, almost all trajectories escape out of the domain after some time
so that the repeller
as well as its stable or unstable manifolds are fractal objects.

These fractals can be generated by allowing the escape of trajectories over
a long but finite
time interval.  Over a finite time, there remain a sizable set of
trajectories, which
progressively reduces to the fractal as the time interval becomes longer
and longer.  The number
of trajectories in the set (or statistical ensemble) decays with time.
Typically, the decay is
exponential and characterized by the so-called escape rate $\gamma$.

The escape rate $\gamma$ can be evaluated by solving the first passage
problem of the Helfand
moment by introducing absorbing boundaries at $\tilde
G_{yx}=\pm\frac{\chi}{2}$.  Indeed, the
Einstein-Helfand equation (\ref{Einstein}) shows that the Helfand moment
performs a
diffusive-like random walk.  Accordingly, the Helfand moment can be
considered as a random
variable $g=\tilde G_{yx}$ for which the probability density $p(g)$ obeys
to a diffusion-type
equation \cite{dorf-gasp}:
\begin{equation}
\frac{\partial p}{\partial t} = \eta \; \frac{\partial ^{2}p}{\partial
g^{2}} \; ,
\label{fokker-planck}
\end{equation}
where the role of the diffusion coefficient is played by the shear viscosity
(\ref{Einstein}) itself.  At the absorbing boundaries, the probability
density must satisfy the
absorbing boundary conditions:
\begin{equation}
p\left(- \frac{\chi }{2}\right) = p\left(+ \frac{\chi }{2}\right) = 0 \; .
\label{abc.p}
\end{equation}
The solution of the diffusion-type equation (\ref{fokker-planck}) with the
boundary conditions
(\ref{abc.p}) is given by
\begin{equation}
p(g,t) = \sum_{n=1}^{\infty } c_{n} \; \exp(-\gamma _{n} t) \; \sin \left
\lbrack
\frac{\pi n}{ \chi }
\left ( g +  \frac{\chi }{2} \right ) \right \rbrack \; ,
\end{equation}
with
\begin{equation}
\gamma _{n}= \eta \left ( \frac{\pi n}{\chi } \right )^{2} \; ,
\end{equation}
and where the coefficient $c_n$ depend on the initial probability density.
The number ${\cal N}(t)$ of trajectories remaining between the absorbing
boundaries at the
current time $t$ is related to the probability density by:
\begin{equation}
{\cal N}(t) = {\cal N}_{0} \int _{-\frac{\chi }{2}}^{+\frac{\chi }{2}} \:
p(g,t) \; dg \sim {\cal
N}_{0} \; \exp(-\gamma_1 t) \; .
\end{equation}
After a long time, the escape is dominated by the smallest decay rate,
$\gamma_{1}$, which can therefore be identified with the escape rate
$\gamma$. In this way, we
obtain the \textit{escape rate} as a function of $\chi$:
\begin{equation}
\gamma =\gamma_{1} = \eta \left ( \frac{\pi}{\chi } \right )^{2} \; .
\label{taux-esc}
\end{equation}
This result is obtained by using the diffusion-type equation
(\ref{fokker-planck}) which is
expected to hold over spatial distances larger than the mean free path of
the particles.
Therefore, the parameter $\chi$ must be sufficiently large so that the
Helfand moment is in a
diffusion regime and Eq. (\ref{fokker-planck}) holds.

The shear viscosity coefficient can thus be obtained from the escape rate
which depends on the
parameter $\chi$ of separation between the absorbing boundaries as
\begin{equation}
\eta = \lim_{\chi\to\infty} \left ( \frac{\chi }{\pi  } \right )^{2}
\gamma(\chi) \; .
\label{visc-esc}
\end{equation}
In the following we call Eq. (\ref{visc-esc}) the escape-transport formula.

\subsection{The chaos-transport formula}

At the microscopic level of description, the escape rate is controlled by
the fractal
repeller ${\cal F}_{\chi}$ which is composed of all the trajectories which
satisfy the condition
(\ref{abc}) under forward and backward time evolution.  The repeller is the
support of a natural
invariant probability measure.  This invariant measure is natural because
it is generated by the
dynamics and can be approximated by a statistics based on the trajectories
remaining within the
absorbing boundaries after a long but finite time. The dynamics on the
fractal repeller is
characterized by positive Lyapunov exponents and a KS entropy, both
evaluated with respect to the
natural invariant measure of the repeller.  If the dynamics is unstable
some Lyapunov exponents
are positive.  If the dynamics is chaotic the KS entropy is positive.  On a
repeller, the sum of
positive Lyapunov exponents differs from the KS entropy and the difference
gives the escape rate
$\gamma(\chi)$ of the repeller ${\cal F}_{\chi}$:
\cite{eckmann-ruelle,kantz-grassberger}
\begin{equation}
\gamma(\chi) = \left(\sum_{\lambda_i>0} \lambda_i - h_{\rm
KS}\right)_{{\cal F}_{\chi}} \; .
\end{equation}
If we combine this result from dynamical systems theory with Eq.
(\ref{visc-esc}), we obtain the
chaos-transport formula (\ref{esc.eq}) for viscosity as originally derived
by Dorfman and Gaspard
\cite{dorf-gasp}.

An equivalent formula can be obtained which involves the partial fractal
dimensions of the
repeller instead of the KS entropy.  Indeed, the fractal character of the
repeller is a direct
consequence of the escape of trajectories so that the KS entropy is no
longer equal to the sum of
Lyapunov exponents but to
\begin{equation}
h_{\rm KS} = \sum_{\lambda_i>0} d_i \; \lambda_i \; ,
\end{equation}
where the coefficients are the partial information dimensions of the
repeller associated with
each unstable direction of corresponding Lyapunov exponent $\lambda_i$
\cite{eckmann-ruelle}.
These partial dimensions satisfy
\begin{equation}
0 \leq d_i \leq 1 \; ,
\end{equation}
so that the KS entropy is in general smaller than the sum of positive
Lyapunov exponents.
Accordingly, the escape rate can be expressed as
\begin{equation}
\gamma(\chi) = \left(\sum_{\lambda_i>0} c_i \; \lambda_i\right)_{{\cal
F}_{\chi}} \; 
\end{equation}
in terms of the partial codimensions defined as
\begin{equation}
c_i \equiv 1 - d_i \; .
\end{equation}
Combining with Eq. (\ref{visc-esc}), the shear viscosity is given by
\begin{equation}
\eta = \lim_{\chi\to\infty} \left ( \frac{\chi }{\pi  } \right )^{2}
\left(\sum_{\lambda_i>0} c_i
\; \lambda_i\right)_{{\cal F}_{\chi}} \; .
\label{visc-codim}
\end{equation}
In the limit $\chi\to\infty$, the Lyapunov exponents reach their
equilibrium values
$\lambda_{i,{\rm eq}}$, while the codimensions vanish typically as $c_i
\sim \chi^{-2}$ if
transport is normal. If we introduce the quantities
\begin{equation}
a_i \equiv \lim_{\chi\to\infty} \left ( \frac{\chi }{\pi  } \right )^{2}
c_i\Big\vert_{{\cal
F}_{\chi}}
\; ,
\label{coeff.a}
\end{equation}
Eq. (\ref{visc-codim}) provides a decomposition of the viscosity
coefficient on the spectrum of
Lyapunov exponents such as
\begin{equation}
\eta = \sum_{\lambda_{i,{\rm eq}}>0} a_i
\; \lambda_{i,{\rm eq}} \; .
\label{decomp}
\end{equation}
Typically, the escape is most important in the most unstable direction
corresponding to the
maximum Lyapunov exponent $\lambda_1$.  Therefore, the repeller is more
depleted in the most unstable direction and the corresponding partial
dimension $d_1$ is lower
than the further ones.  This reasoning suggests that a typical behavior is
\begin{equation}
\left(\sum_{\lambda_i>0} c_i \; \lambda_i\right)_{{\cal F}_{\chi}} \simeq
\left(c_1 \; \lambda_1\right)_{{\cal F}_{\chi}}\; ,
\label{typic}
\end{equation}
for $\chi\to\infty$ if the maximum Lyapunov exponent $\lambda_1$ is well
defined.

This is precisely the case in two-degree-of-freedom systems such as the
two-disk model where the
chaos-transport formula reduces to
\begin{equation}
\eta = \lim_{\chi\to\infty} \left ( \frac{\chi }{\pi  } \right )^{2} \left(
c_{\rm I}
\; \lambda\right)_{{\cal F}_{\chi}} \; ,
\label{visc-codim2}
\end{equation}
where $\lambda$ is the unique positive mean Lyapunov exponent and $c_{\rm
I}$ the corresponding
codimension which should be understood as the partial information
codimension of the
unstable manifolds of the fractal repeller given in terms of the partial
information dimension by
Young's formula \cite{young}
\begin{equation}
c_{\rm I} = 1-d_{\rm I} = 1 - \frac{h_{\rm KS}}{\lambda} \; .
\end{equation}
It is known that the partial information dimension of the repeller is well
approximated by the
partial Hausdorff dimension if the escape rate is small enough and if
Ruelle's topological
pressure does not present a discontinuity.  This last condition is
fulfiled if the system does
not undergo a dynamical phase transition, which is the case in the
finite-horizon regimes of
Sinai's billiard which controls the dynamics of both the Lorentz gas and
the two-disk
model \cite{gasp-baras}.  Under these conditions, we can replace the
partial information
codimension by the partial Hausdorff codimension in the chaos-transport
formula and obtain the
viscosity as
\begin{equation}
\eta = \lim_{\chi\to\infty} \left ( \frac{\chi }{\pi  } \right )^{2} \left(
c_{\rm H}
\; \lambda \right)_{{\cal F}_{\chi}} \; .
\label{visc-Hcodim}
\end{equation}
In the limit $\chi\to\infty$, the Lyapunov exponent converges to its
equilibrium value so that we
can also write:
\begin{equation}
\eta = \lambda_{\rm eq} \; \lim_{\chi\to\infty} \left ( \frac{\chi }{\pi  }
\right )^{2}
c_{\rm H}(\chi) \; ,
\label{visc-Hcodim-lambda_eq}
\end{equation}
under similar conditions as Eq. (\ref{visc-Hcodim}).


\section{Fractal repeller}
\label{repeller}

In this section, our purpose is to display the fractal repeller associated
with viscosity in the
two-disk model and to compare it with the fractal repeller of diffusion in
the Lorentz gas in
order to show that they are different and therefore specific to each
transport property.

\subsection{Shear viscosity in the two-disk model}

The two-disk model has been studied by Bunimovich and Spohn who proved
thanks to a central limit
theorem that viscosity is positive and finite in this minimal model
\cite{spohn-buni}.  In the
previous paper \cite{premier-article}, we have considered the two-disk
model in the hexagonal
geometry which we shall use in the following.  We showed in Ref.
\cite{premier-article} that the
dynamics reduces to a Sinai billiard in the center-of-mass frame and that
the Helfand moment
(\ref{Helfand}) with $N=2$ is then given by
\begin{equation}
\tilde G_{yx}(t) = \sqrt{\frac{\beta}{V}} \left[
x(t) \; p_{y}(t) - \sum_s \Delta x^{(s)} \; p_{y}^{(s)} \; \theta(t-t_s)
\right] \; ,
\label{Helfand2}
\end{equation}
where $(x,y)$ are the coordinates of the relative position of both disks
and $(p_x,p_y)$ the
canonically conjugated relative momentum.  The jumps happen when the
trajectory of Sinai's
billiard crosses the hexagonal boundary.  If the trajectory crosses the
side of label
$\omega$ the trajectory is reinjected at the opposite side so that the jump
in position is given
by the lattice vector $\Delta{\bf r}^{(s)}=-{\bf c}_{\omega}^{(s)}$
corresponding to the side
$\omega$.

\begin{figure}[h]
\centerline{\epsfig{file=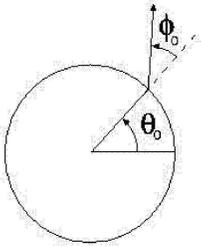,width=3cm}}
\vspace*{0.2cm}
\caption{Initial conditions of the particules in the Sinai billiard.}
\label{variables-particules}
\end{figure}

A fractal repeller is defined by considering all the trajectories such that
their Helfand moment
satisfies the conditions
\begin{equation}
- \frac{\chi}{2} \leq \tilde G_{yx} \leq + \frac{\chi}{2} \; ,
\label{abc-visc}
\end{equation}
where the parameter $\chi$ should be large enough.  The stable manifolds of
the fractal repeller
can be visualized by plotting the initial conditions of trajectories
satisfying the conditions
(\ref{abc-visc}) over a long time interval extending forward in time.
These initial conditions
are taken on the disk of Sinai's billiard.  The initial conditions are
specified by the angle
$\theta$ of the initial position and the angle $\phi$ that the initial
velocity makes with a
vector which is normal to the disk at the initial position (see Fig.
\ref{variables-particules}).
The initial conditions are plotted in the Birkhoff coordinates
($\theta$,$\sin\phi$).

\begin{figure}[h]
\centerline{\epsfig{file=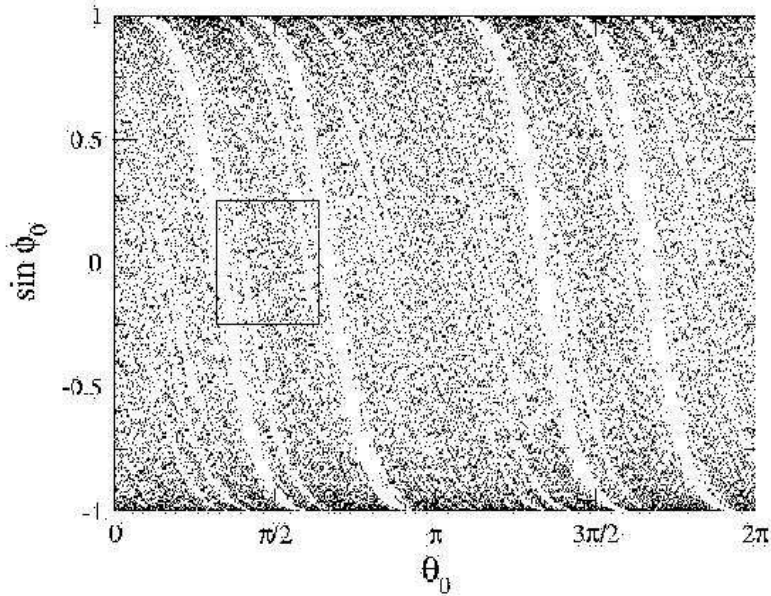,width=12cm}}
\vspace*{0.2cm}
\caption{Fractal repeller associated with viscosity in the hexagonal
geometry with absorbing
boundaries at $\chi=2.70$. The density is $n=(2/V)=0.45$.}
\label{fractal-visc-TOT}
\end{figure}

\begin{figure}[h]
\centerline{\epsfig{file=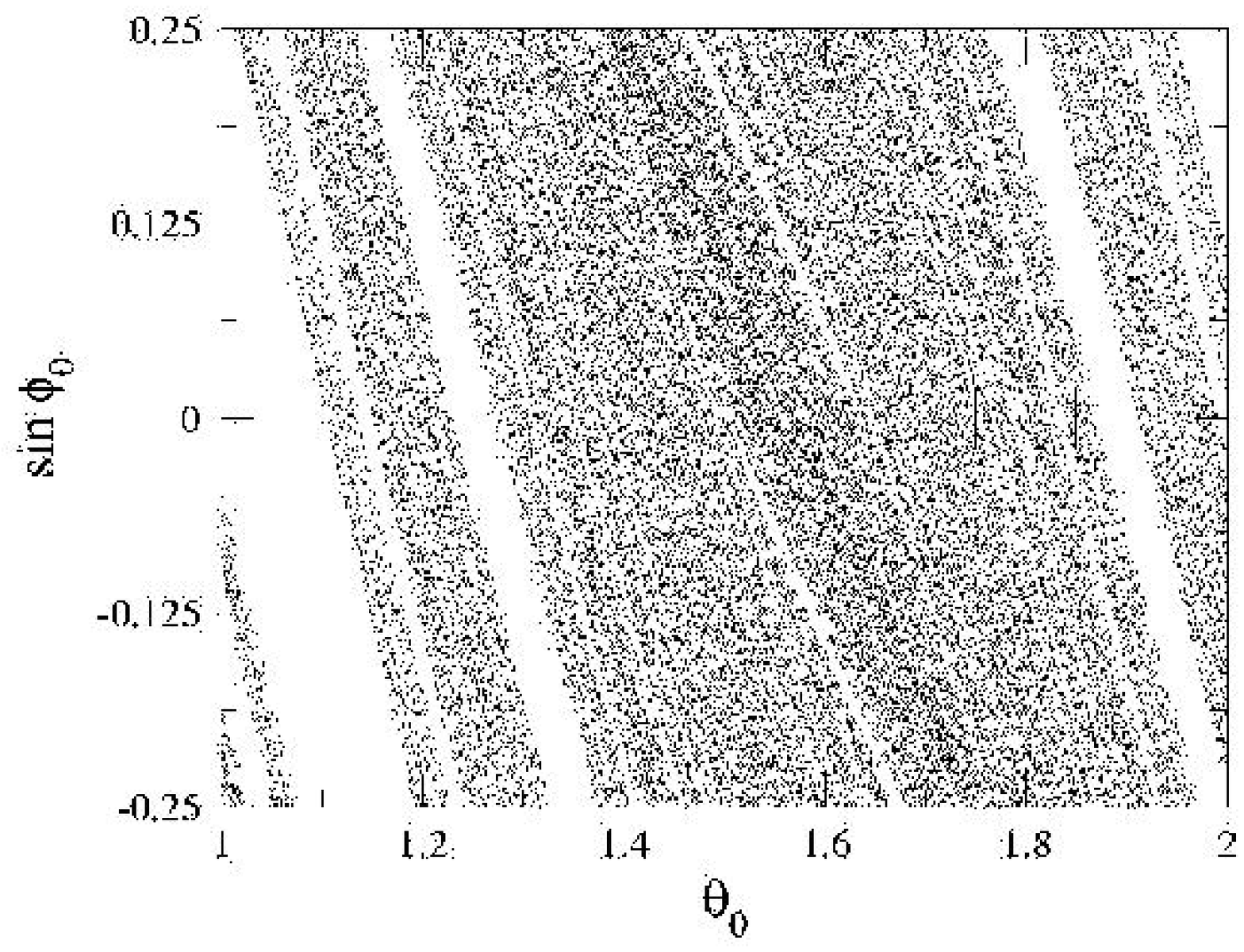,width=12cm}}
\vspace*{0.2cm}
\caption{Enlarging of the domain in
to the square in Fig. \ref{fractal-visc-TOT}.}
\label{fractal-visc-zoom1}
\end{figure}

\begin{figure}[h]
\centerline{\epsfig{file=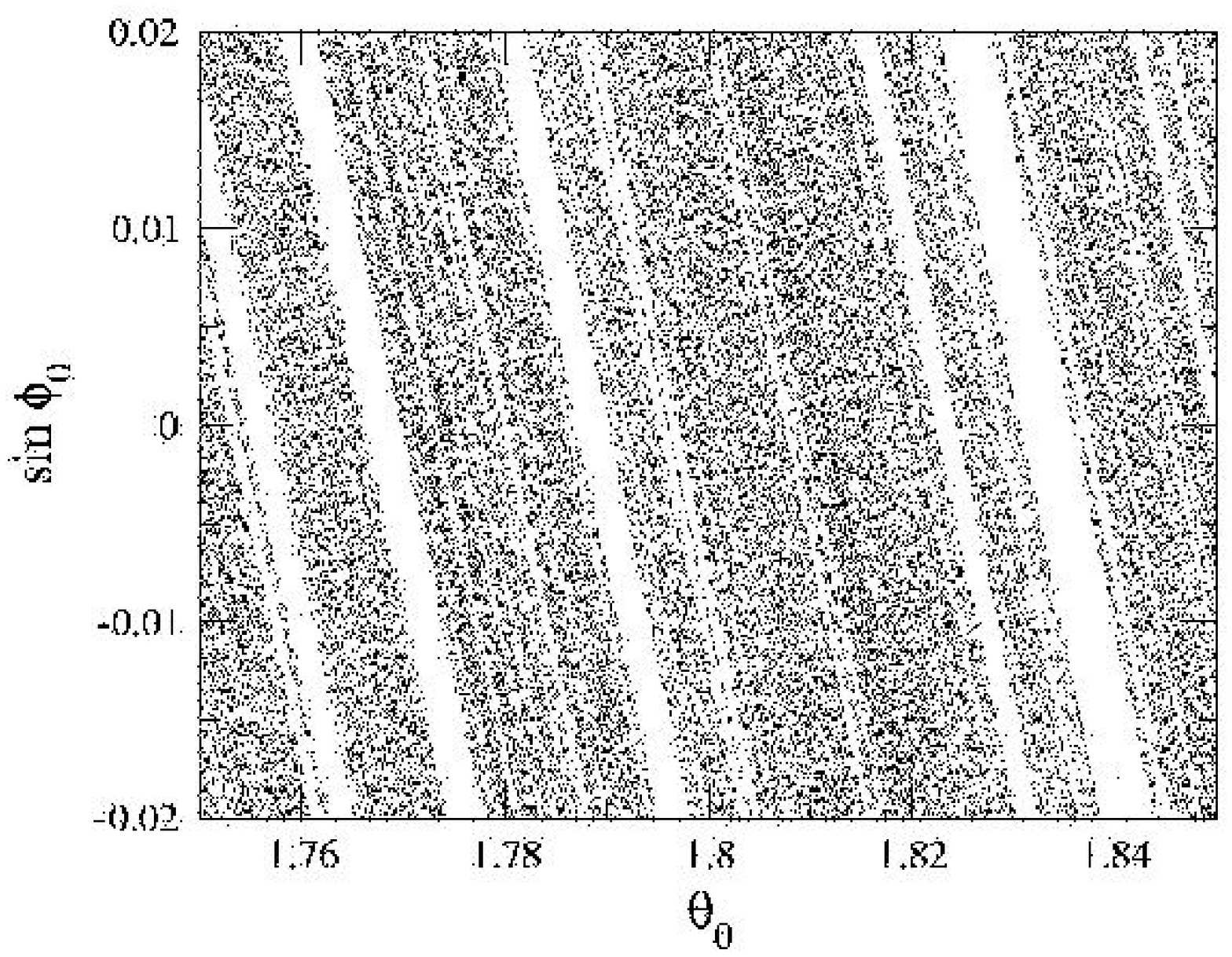,width=12cm}}
\vspace*{0.2cm}
\caption{Enlarging of the domain into the square in Fig.
\ref{fractal-visc-zoom1}.}
\label{fractal-visc-zoom2}
\end{figure}

Figure \ref{fractal-visc-TOT} depicts such the fractal composed of the
stable manifolds of
the repeller for viscosity in the two-disk model.  We provide evidence that
the set is
fractal by zooming successively on it in Figs.
\ref{fractal-visc-zoom1} and
\ref{fractal-visc-zoom2}, where the self-similarity of the repeller
clearly appears.

\begin{figure}[h]
\centerline{\epsfig{file=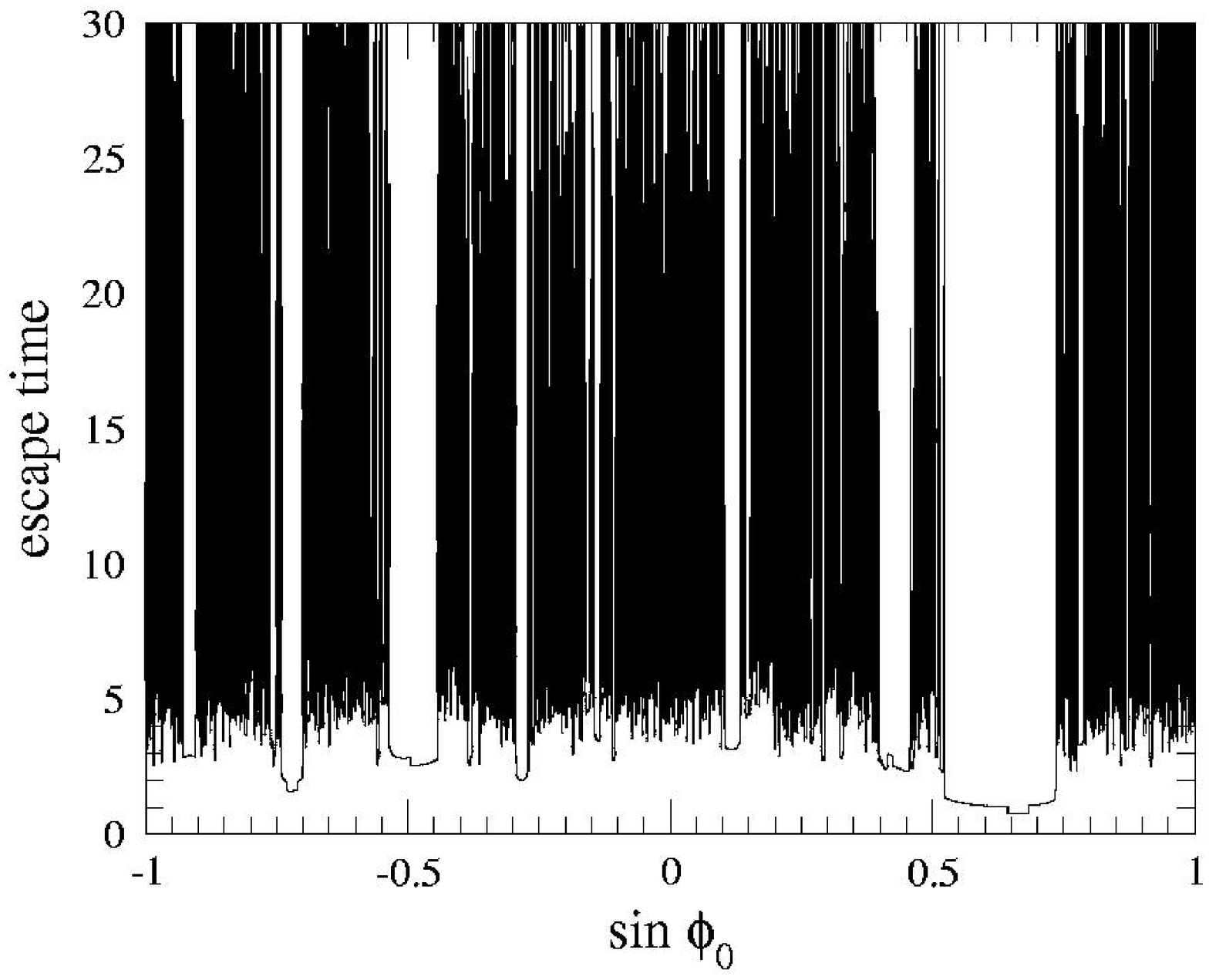,width=10cm}}
\vspace*{0.2cm}
\caption{Escape-time function versus $\sin \phi_{0}$ ($\theta_{0} = \pi/4$).
This function corresponds to a section in Fig. \ref{fractal-visc-TOT} along a
vertical line at $\theta_{0} = \pi/4$.}
\label{temps-esc-visc-8G}
\end{figure}

Let us take a section across the repeller in Fig. \ref{fractal-visc-TOT} at
$\theta_{0}=\pi/4$. Taking the
escape time of the corresponding trajectory, we have obtained the escape-time
function depicted in Fig. \ref{temps-esc-visc-8G}.  The time for the
trajectory to escape out
of the phase-space region corresponding to the interval (\ref{abc-visc}) is
infinite if the
trajectory belongs to the stable manifold of a trajectory of the repeller.
Indeed, this
trajectory is then asymptotic to a trajectory which does not escape.
Accordingly, the
escape-time function has vertical asymptotes on the stable manifolds of the
repeller.  Since
the repeller is fractal the vertical asymptotes are not enumerable, which
explains the
behavior in Fig. \ref{temps-esc-visc-8G}.

\vspace*{0.5cm}

\vspace*{1cm}

\subsection{Diffusion in the Lorentz gas}

Diffusion of a tracer particle in the hard-disk periodic Lorentz gas has
been studied with the
escape-rate formalism in Ref. \cite{gasp-baras}.  In this Lorentz gas, the
tracer particle
undergoes elastic collisions on hard disks forming a triangular lattice.
In a unit cell of the
lattice, the dynamics also reduces to Sinai's billiard.  The energy of the
tracer particle is
conserved as well as the phase-space volumes.  Sinai and Bunimovich have
proved that the
dynamics is ergodic and mixing and that the diffusion coefficient is
positive and finite in the
finite-horizon regime \cite{sinai-buni}.  For diffusion, the associated
Helfand moment is simply
given by one of the coordinates $(x,y)$ of position of the tracer particle
\cite{dorf-gasp}.  An
escape process is associated with diffusion by setting up a problem of
first passage of the tracer
particle at some absorbing boundaries.  If we consider the $x$-coordinates,
the tracer particle
does not escape as long as the following condition is satisfied:
\begin{equation}
-\frac{R}{2}\leq x \leq + \frac{R}{2} \; . \label{abc-diff}
\end{equation}
The absorbing boundary conditions are therefore defined at $x=\pm\frac{R}{2}$.
With these absorbing boundaries, the system is called an \textit{open
Lorentz gas}
\cite{gasp-baras}.

The trajectories trapped within the interval (\ref{abc-diff}) form a
fractal repeller as shown in
Ref. \cite{gasp-baras}.  In order to compare with the fractal repeller of
viscosity, we can plot
the fractal repeller of diffusion in a similar way as here above for viscosity.

Here again, we plot all the initial conditions of trajectories remaining
within the interval
(\ref{abc-diff}) over a long forward time interval.  These initial
conditions are plotted in the
same Birkhoff coordinates ($\theta$,$\sin\phi$) of a disk around the
coordinate $x\simeq 0$ in the
Lorentz gas.  The set of the selected initial conditions approximate the
stable manifolds of the
fractal repeller.  We zoom successively on this fractal in Figs.
\ref{fractal-diff-zoom1} and
\ref{fractal-diff-zoom2}, which provides evidence of its self-similarity.
As a consequence, the
repeller is also fractal.  The fractal dimension of the repeller is related
to the diffusion
coefficient of the Lorentz gas and its Lyapunov exponent, as shown in Refs.
\cite{gasp-nicolis,gasp-baras}.

\vspace*{1cm}
\begin{figure}[h]
\centerline{\epsfig{file=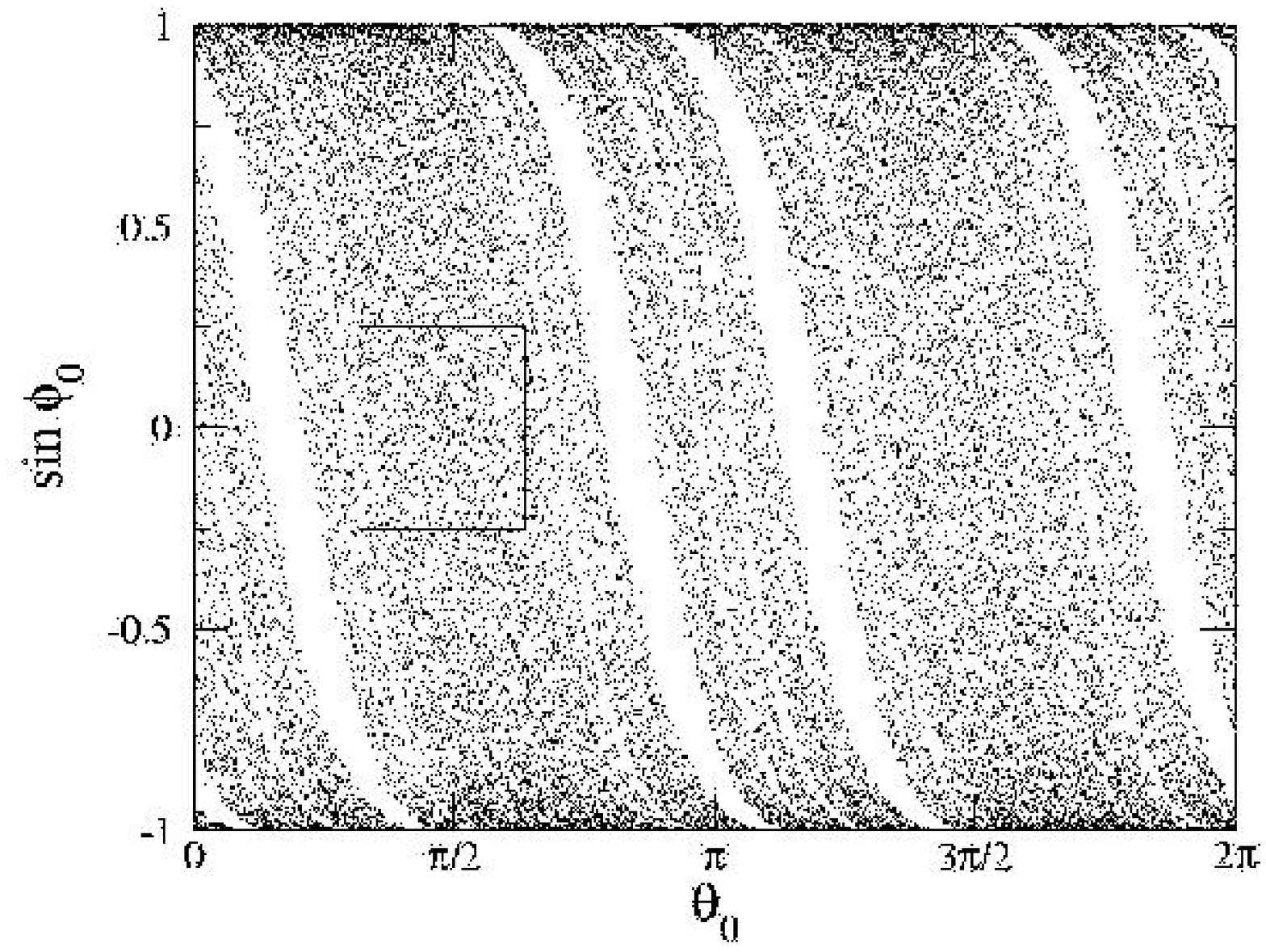,width=12cm}}
\vspace*{0.2cm}
\caption{Fractal repeller associated with diffusion in the hexagonal
geometry with absorbing
boundaries at $R=4$.  The density of hard disks is $n=0.45$.}
\label{fractal-diff-TOT}
\end{figure}

\begin{figure}[h]
\centerline{\epsfig{file=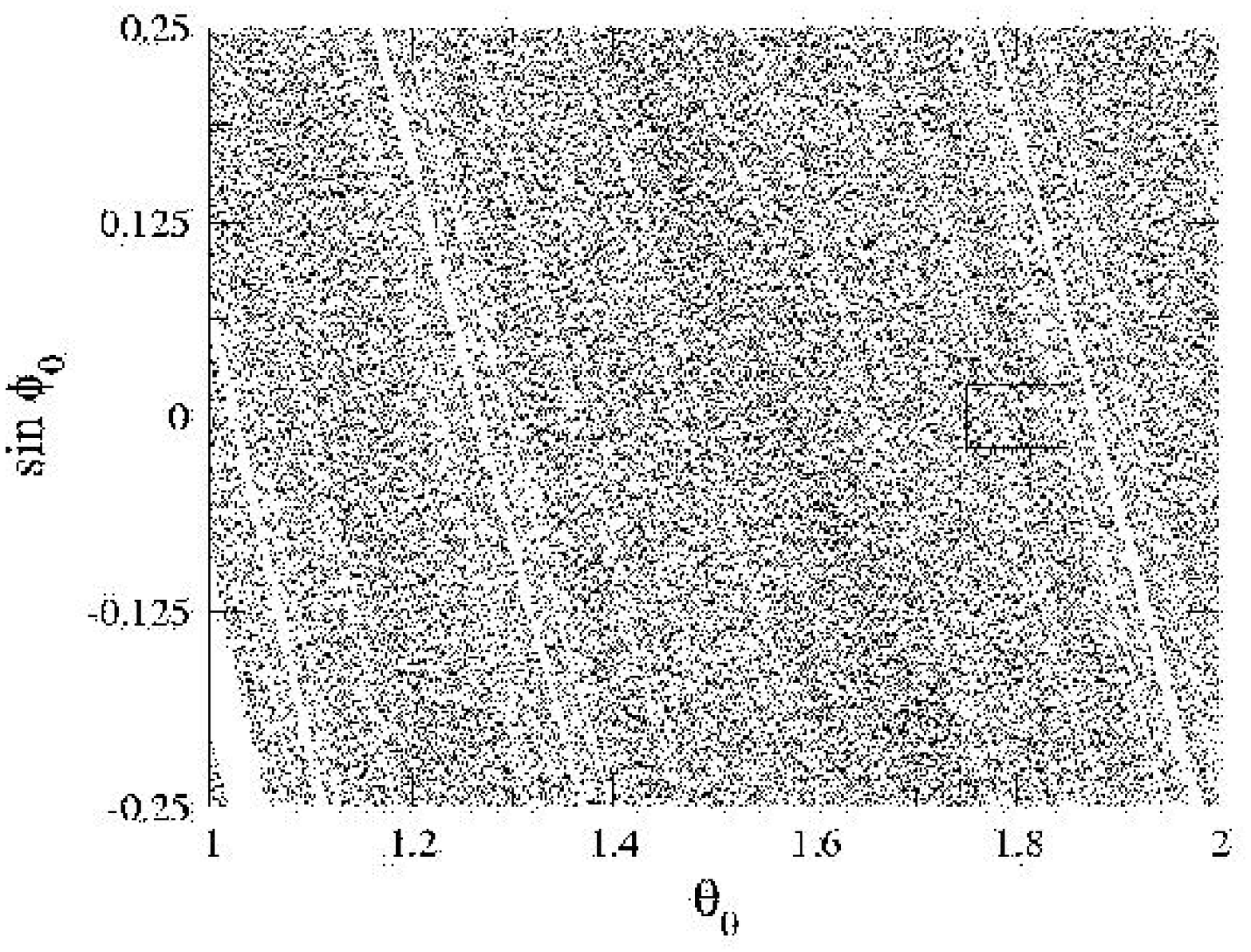,width=12cm}}
\vspace*{0.2cm}
\caption{Enlarging of the domain into the square in Fig.
\ref{fractal-diff-TOT}.}
\label{fractal-diff-zoom1}
\end{figure}

\begin{figure}[h]
\centerline{\epsfig{file=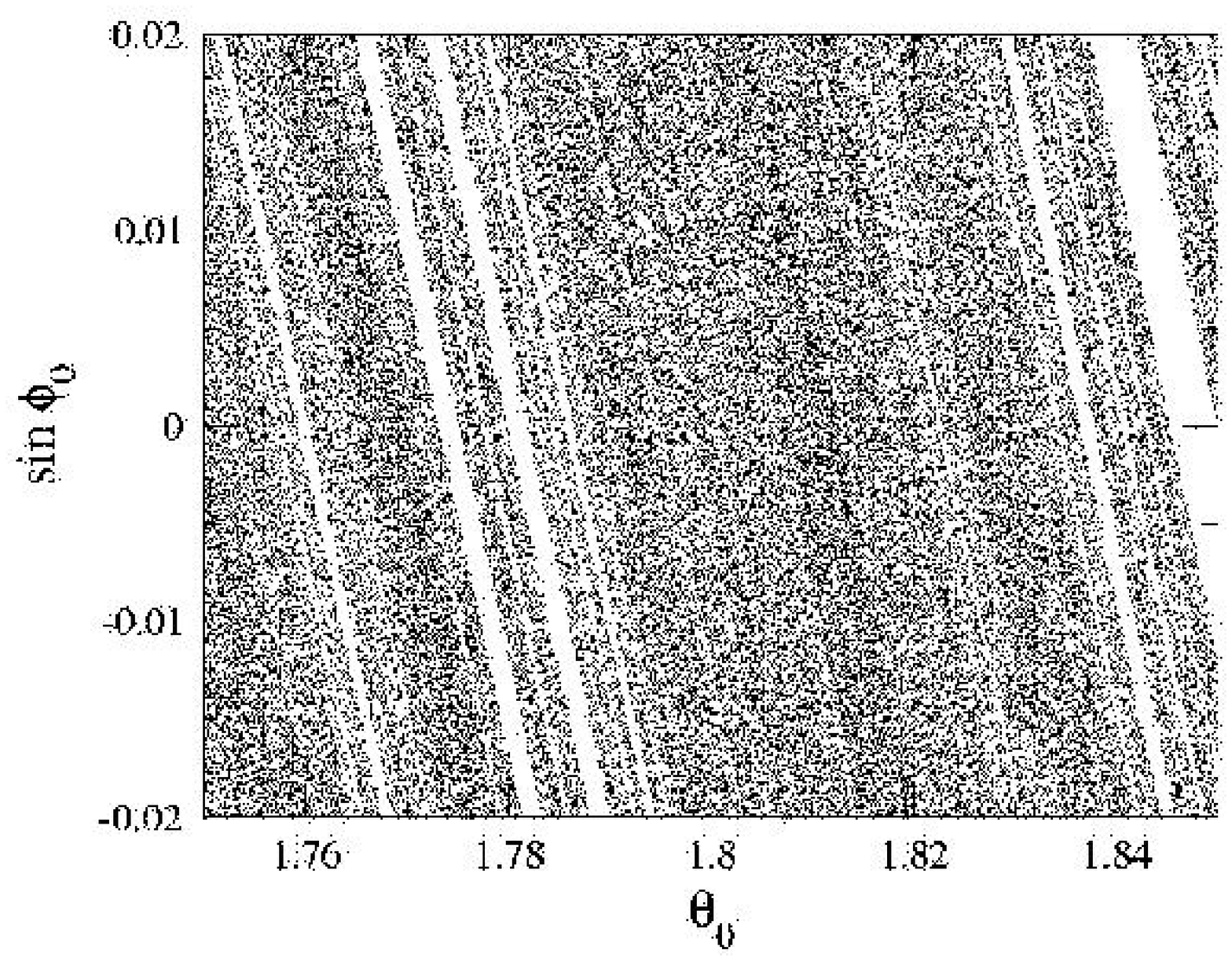,width=12cm}}
\vspace*{0.2cm}
\caption{Enlarging of the domain into the square in Fig.
\ref{fractal-diff-zoom1}.}
\label{fractal-diff-zoom2}
\end{figure}

\subsection{Comparison between diffusion and viscosity}

We observe that the two fractal repellers associated respectively with
diffusion (see
Fig. \ref{fractal-diff-TOT}) and viscosity (see Fig. \ref{fractal-visc-TOT}) are
different. Indeed, although the global structure is similar, the trajectories
belonging to the different repellers are not the same.

\begin{figure}[h]
\centerline{\epsfig{file=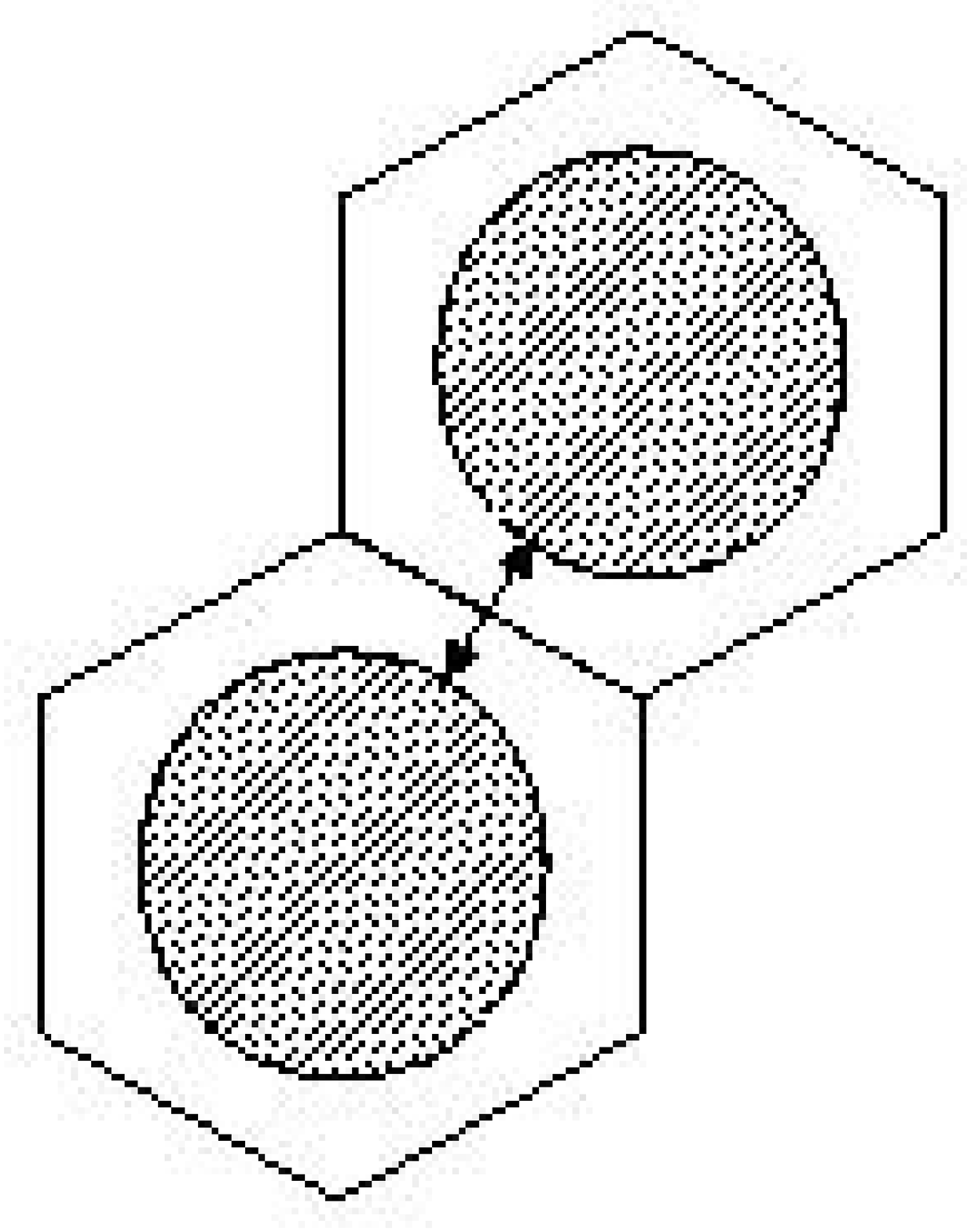,width=3cm}}
\vspace*{0.2cm}
\caption{Periodic trajectory belonging to the fractal repeller associated with
diffusion but not to the one associated with viscosity.}
\label{repeller-comparaison1}
\end{figure}

To convince us of this difference, we take some examples of trajectories. In
Fig.
\ref{repeller-comparaison1}, we have a periodic trajectory bouncing between
two disks in the
billiard.  This trajectory belongs to the repeller associated with
diffusion since the position
$x$ is bounded and satisfied (\ref{abc-diff}).  However, the viscosity
Helfand moment of this
trajectory does not satisfy the condition (\ref{abc-visc}) so that it does
not belong to the
repeller of shear viscosity. With Eq.
(\ref{Helfand2}), we see that, in one direction, both
$\Delta x^{(s)}$ and $p_{y}^{(s)}$ are positive. Therefore, the
contribution at this passage is
positive for the Helfand moment. In the other direction, both
$\Delta x^{(s)}$ and $p_{y}^{(s)}$ are negative but the product  $\Delta
x^{(s)} p_{y}^{(s)}$ is
also positive. Accordingly, the Helfand moment increases forever on this
trajectory which,
therefore, does not belong to the repeller associated with shear viscosity.

\begin{figure}[h]
\centerline{\epsfig{file=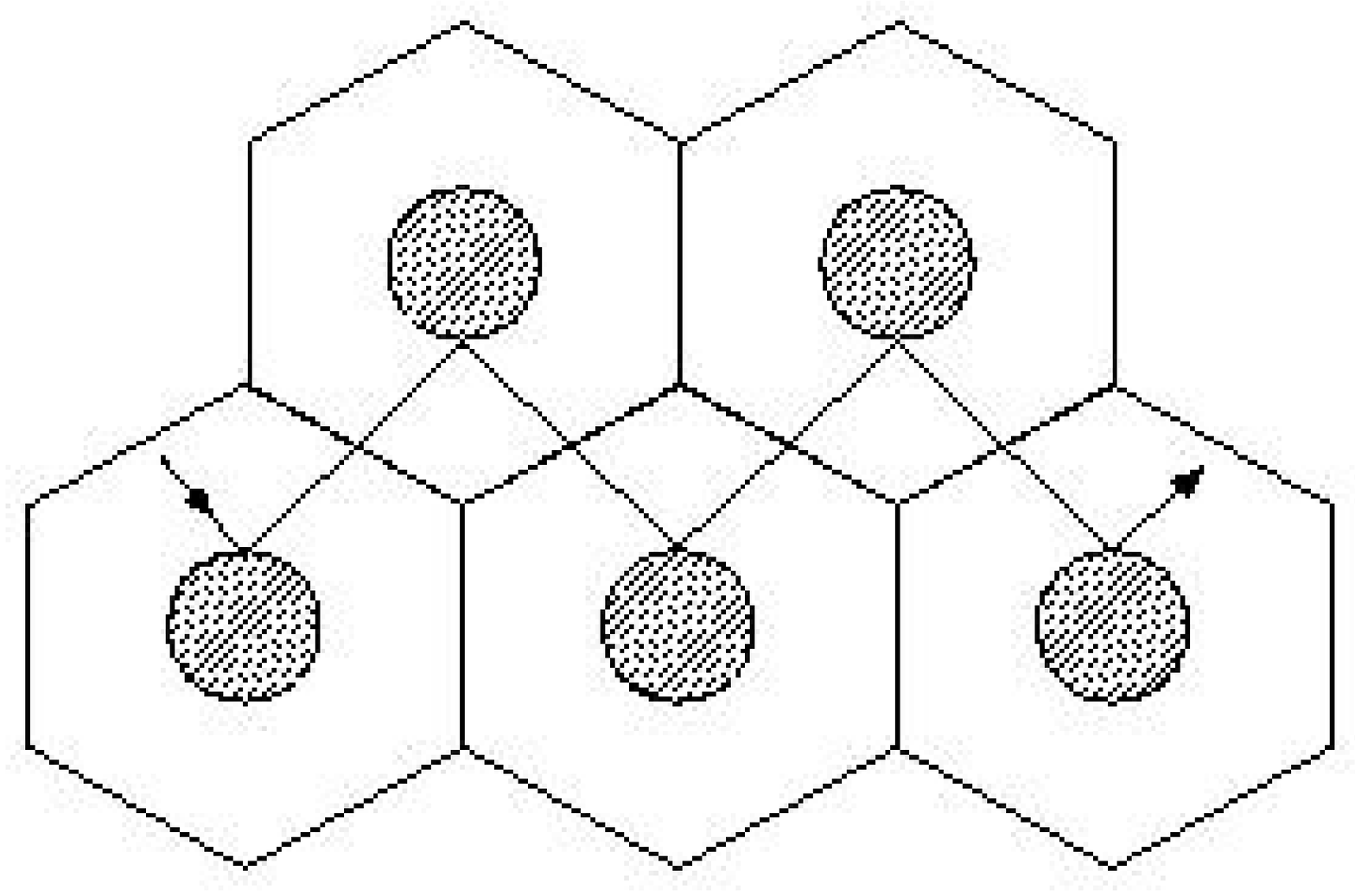,width=6cm}}
\vspace*{0.2cm}
\caption{Typical trajectory which moves through the whole system but which
has a Helfand moment that remains close to zero.  This trajectory belongs to the
repeller of viscosity but not to the one of diffusion.}
\label{exemple-diff-non-repeller}
\end{figure}

On the other hand, we can observe the opposite case. Figure
\ref{exemple-diff-non-repeller} depicts an example of trajectory escaping
from the
interval (\ref{abc-diff}) although its Helfand moment of viscosity remains
between the absorbing
boundary conditions (\ref{abc-visc}).

The repellers associated with different transport properties are therefore
different.

\subsection{Escape rate and viscosity}

In this subsection, we show that the shear viscosity can be obtained from
the escape rate of the
repeller by using the escape-transport formula (\ref{visc-esc}).  We
consider a sequence of
repellers with larger and larger values of the parameter $\chi$.  The
escape rate $\gamma(\chi)$
is numerically evaluated for each repeller by computing the decay of the
number ${\cal N}(t)$ of
trajectories still within the interval (\ref{abc-visc}) at current time and
by extracting the
escape rate from the exponential decay.  The escape rate is observed to
behave as $\gamma(\chi)
\sim \chi^{-2}$ and the shear viscosity coefficient is then obtained with
Eq. (\ref{visc-esc}).

\begin{figure}[!h]
\centerline{\epsfig{file=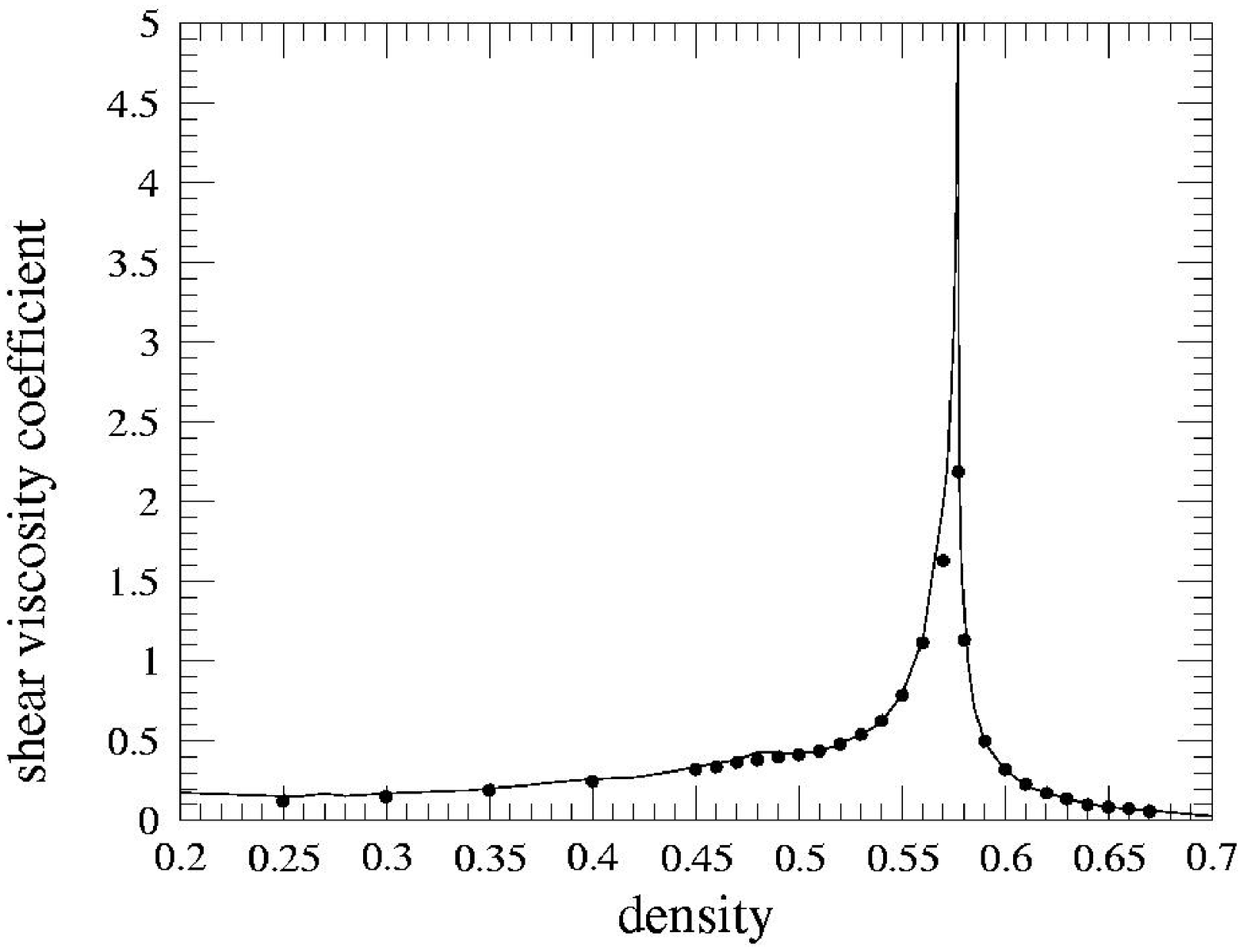,width=10cm}}
\vspace*{0.2cm}
\caption{Comparison between two methods of calculating the shear viscosity
coefficient $\eta^*=\eta_{xy,xy}^*$ in the hexagonal geometry: the Einstein-Helfand formula
(continuous line) and the escape-transport formula (\ref{visc-esc}) with
$\chi=60\sqrt{n}$ (dots).}
\label{superposition-visc-helf-taux-hex}
\end{figure}

\begin{figure}[!h]
\centerline{\epsfig{file=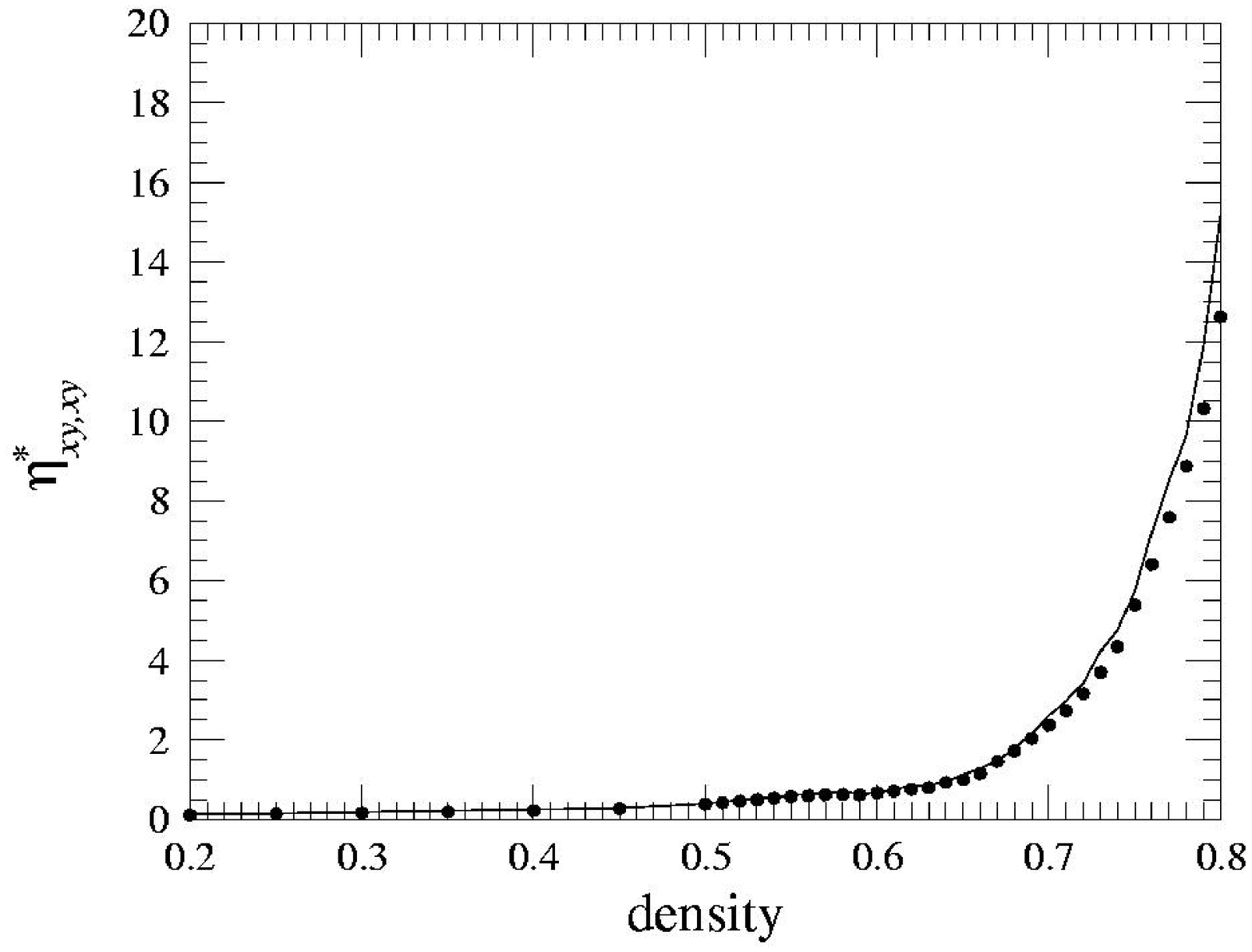,width=10cm}}
\vspace*{0.2cm}
\caption{Comparison between two methods of calculating the shear viscosity
coefficient $\eta_{xy,xy}^*$ in the square geometry: the Einstein-Helfand formula
(continuous line) and the escape-transport formula (\ref{visc-esc})
with $\chi=45\sqrt{n}$ for density $n<0.66$, $\chi=100\sqrt{n}$ for $0.67<n<0.75$, and
$\chi=150\sqrt{n}$ for $0.76<n$ (dots).}
\label{superposition-visc-helf-taux-carre}
\end{figure}

Figures \ref{superposition-visc-helf-taux-hex} and
\ref{superposition-visc-helf-taux-carre} depict the viscosity directly
computed from the escape
rate and compared with the values obtained by the Einstein-Helfand formula
in Ref.
\cite{premier-article}, respectively in the hexagonal and square
geometries.  As in the previous
paper \cite{premier-article}, we consider reduced viscosities defined by
\begin{equation}
\eta_{ij,kl}^{*} \equiv \frac{\eta_{ij,kl}}{2\sqrt{mk_{\rm B}T} \;} \; .
\end{equation}
We observe in Figs. \ref{superposition-visc-helf-taux-hex} and
\ref{superposition-visc-helf-taux-carre} the excellent agreement between
both methods.

\vspace*{0.5cm}

\section{Viscosity from the chaotic and fractal properties of the repeller}
\label{chaos}

In this section, we compute the shear viscosity coefficient in terms of the
chaotic and fractal
properties of the repeller by using the chaos-transport formula
(\ref{visc-Hcodim}) which related
the viscosity to the Lyapunov exponent and the Hausdorff codimension of the
repeller of viscosity.

\subsection{Lyapunov exponent}

In Sinai's billiard which controls the reduced dynamics of the two-disk
model, the elastic
collisions between the disks are defocusing.  This induces a dynamical
instability of the
trajectories which is characterized by the Lyapunov exponents. These
exponents are the rates of
exponential separations between a reference orbit and infinitesimally close
orbits. Since the
dynamics of Sinai's billiard is symplectic and volume-preserving in the
four-dimensional phase
space, the Lyapunov exponents spectrum is ($+\lambda,0,0,-\lambda$) so that
their sum is
vanishing. One of the Lyapunov exponents vanishes because of the absence of
exponential
separation in the direction of the flow. Another one corresponding to the
direction perpendicular
to the energy shell equals zero because of energy conservation.

\begin{figure}[h]
\centerline{\epsfig{file=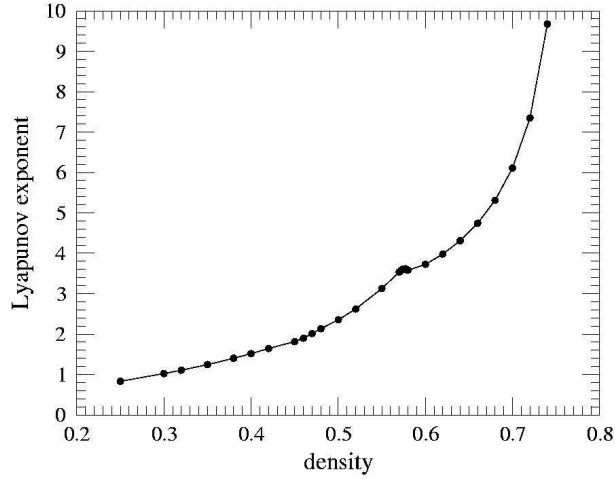,width=8cm}}
\vspace*{0.2cm}
\caption{Equilibrium Lyapunov exponent versus density in the hexagonal
geometry.}
\label{lyap-stretching-hex}
\end{figure}

\begin{figure}[h]
\centerline{\epsfig{file=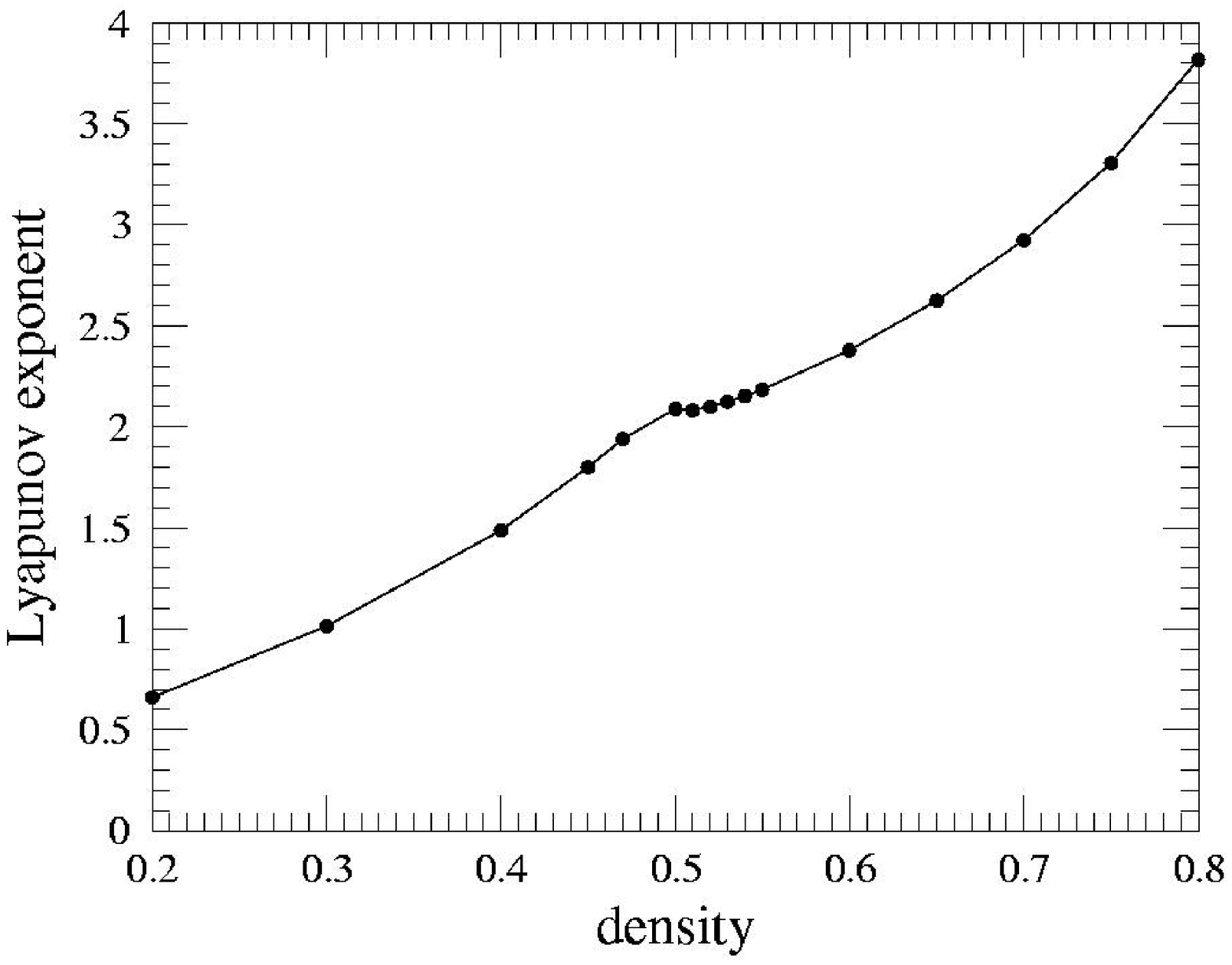,width=8cm}}
\vspace*{0.2cm}
\caption{Equilibrium Lyapunov exponent versus density in the square geometry.}
\label{lyap-stretching-carre}
\end{figure}

There exists a method to calculate the positive Lyapunov exponent by
considering the motion of a front of particles accompanying the reference
particle and issued
from the same initial position but with different initial velocities
\cite{sinai}. Because the dynamics is defocusing, this front is expanding.
Locally on the reference orbit $\Gamma_{t}$ the front (called the unstable
horocycle) is
characterized by a curvature $\kappa_{u}(\Gamma_{t})$ or, equivalently, by
its radius of curvature
$1/\kappa_{u}(\Gamma_{t})$. Thanks to this method explained in detail in
Ref. \cite{gasp-book,gasp-baras}, we have computed the positive Lyapunov
exponent as a
function of the density of the system (in the hexagonal and square
geometries). The equilibrium
values of the Lyapunov exponent are obtained by running a trajectory in
Sinai's billiard
without absorbing boundaries and by averaging over a long time interval.
The resulting numerical
values are depicted in Figs. \ref{lyap-stretching-hex} and
\ref{lyap-stretching-carre}.

In the chaos-transport formula (\ref{visc-Hcodim}), the Lyapunov exponent
has to be evaluated
for the trajectories belonging to the fractal repeller.  The statistical
average is here carried
out for the natural invariant probability measure concentrated on the
fractal repeller.  This
invariant measure defines a nonequilibrium state for the motion.  As
aforementioned, the natural invariant measure is generated by the dynamics
itself.  Accordingly,
the Lyapunov exponent is numerically computed by averaging over a
statistical ensemble of
trajectories which has not yet escaped after a long but finite time.  This
ensemble can be as
large as wished by increasing the number of initial conditions.  In this
way, we can calculate
the nonequilibrium values of the Lyapunov exponent.

In Table I, we present a comparison between the equilibrium
Lyapunov exponent
$\lambda_{\rm eq}$ without absorbing boundary conditions (as depicted in Fig.
\ref{lyap-stretching-hex} and \ref{lyap-stretching-carre} and the
nonequilibrium Lyapunov
exponent $\lambda_{\rm neq}(\chi)$ evaluated over a nonequilibrium measure
which has the fractal
repeller as support. The difference between these exponents is small and 
of the order of the escape rate, in agreement with the results of 
Ref. \cite{vanbei-latz-dorf} for the disordered Lorentz gas.

\vskip 0.8 cm
\hrule height1pt \vskip1pt \hrule
\vskip 0.3 cm

\noindent{Table I. Values of the characteristic quantities of chaos for different
densities $n$ in the hexagonal system: $\lambda_{\rm eq}$ is the equilibrium
Lyapunov exponent for the closed system.  The following quantities characterize the
fractal repeller for viscosity with $\chi=60\sqrt{n}$: $\lambda_{\rm neq}$
is the nonequilibrium Lyapunov exponent of the repeller, $h_{\rm KS}$ its KS entropy, $\gamma$ its
escape rate, $c_{\rm I}$ its partial information codimension, and $c_{\rm H}$ its partial Hausdorff
codimension.}

\vskip 0.5 cm
\hrule
\vskip 0.2 cm
$$
\vcenter{\openup1\jot
\halign{#\hfil&\qquad#\hfil&\qquad#\hfil&\qquad#\hfil&\qquad#\hfil&\qquad#\hfil&\qquad#\hfil\cr
$n$ & $\lambda_{\rm eq}$ & $\lambda_{\rm neq}$ & $\gamma$ & $h_{\rm KS}=\lambda_{\rm neq}-\gamma$
& $c_{\rm I}=\gamma/\lambda_{\rm neq}$ & $c_{\rm H}$\cr
0.40      & 1.5156       & 1.5163  & 0.0017 &  1.5146  &  0.0011  & 0.0011  \cr
0.50      & 2.3519       & 2.3539  & 0.0023 &  2.3516  &  0.00098 & 0.00092 \cr
0.60      & 3.7258       & 3.7249  & 0.0015 &  3.7234  &  0.00040 & \quad - \cr}}$$
\vskip 0.05 cm
\hrule \vskip1pt \hrule height1pt
\vskip 1 cm

\subsection{Hausdorff dimension and viscosity}

In order to determine the viscosity by the chaos-transport formula
(\ref{visc-Hcodim}), we need to
determine the partial Hausdorff codimension $c_{\rm H}$ of the fractal
repeller.  The
corresponding dimension $d_{\rm H}=1-c_{\rm H}$ is the Hausdorff dimension
of the vertical
asymptotes of the escape-time function depicted in Fig.
\ref{temps-esc-visc-8G}.  Its values range in the interval $0\leq d_{\rm
H}\leq 1$.

The Hausdorff codimension can be obtained using the following numerical
algorithm
developed by the group of Maryland \cite{maryland-group}. We consider an
ensemble of pairs of
trajectories starting from initial conditions $\phi_{0}$ differing in a
value $\epsilon$.  The
time taken by the trajectories to escape out of the system is given by the
escape-time function in Fig. \ref{temps-esc-visc-8G}.  The pair is said to be
\textit{uncertain} if the trajectories and their
Helfand moment present at least one of the following conditions: (i) both
trajectories follow
paths that differ by the successive passages through the cell boundaries,
that is, if we
associate to each trajectory a symbolic sequence
($\omega_{1},\omega_{2},...$) which gives the
labels of the cell boundaries across which the successive passages occur,
and both
sequences are different; (ii) one of both trajectories has its Helfand
moment which
reaches the fixed absorbing boundaries (\ref{abc-visc}) when the Helfand
moment of the other one still remains within these limits. If the pair does
not present
one of these conditions it is called \textit{certain}. The fraction
$f(\epsilon)$ of
uncertain pairs in the initial ensemble is known to behave as the power
\begin{equation}
f(\epsilon) \sim \epsilon^{c_{\rm H}} \; ,
\end{equation}
giving the Hausdorff codimension as its exponent.
Derivations of this result can be found elsewhere
\cite{ott,claus-gasp-vanbei,maryland-group}.
This method has been already used in Refs.
\cite{gasp-baras,claus-gasp,claus-gasp-vanbei}.  

We have here applied the Maryland algorithm to obtain the Hausdorff codimension
of the fractal repeller of viscosity.  Table I compares
the partial Hausdorff codimension with the partial
information dimension in particular cases.  We observe that both
codimensions take very close values as expected. 

\begin{figure}[!h]
\centerline{\epsfig{file=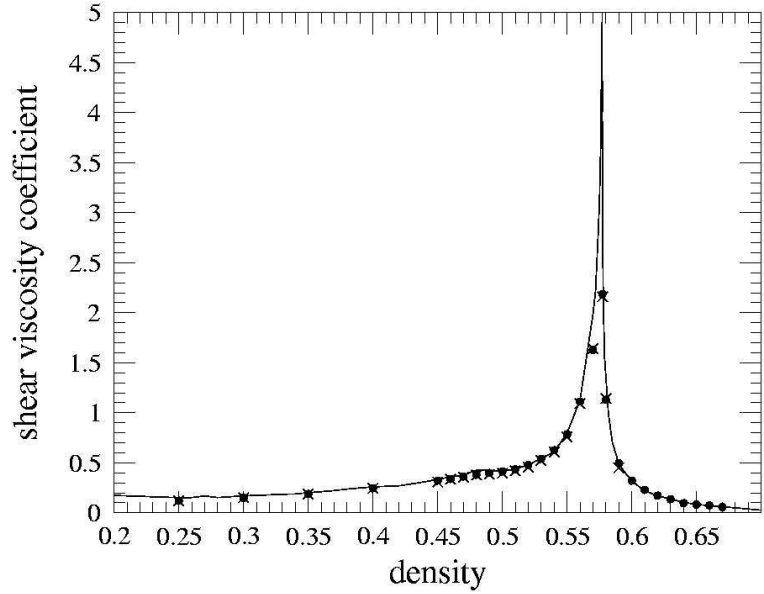,width=10cm}}
\vspace*{0.2cm}
\caption{Comparison between the three methods calculating the shear viscosity
coefficient $\eta^*$ in the hexagonal geometry: the Einstein-Helfand formula
(\ref{Helfand}) (continuous line), the escape-transport formula (\ref{visc-esc})
(dots), and the chaos-transport formula (\ref{visc-Hcodim}) (crosses) with
$\chi=60\sqrt{n}$.}
\label{visc-helf-taux-dim-hex_dens}
\end{figure}

\begin{figure}[!h]
\centerline{\epsfig{file=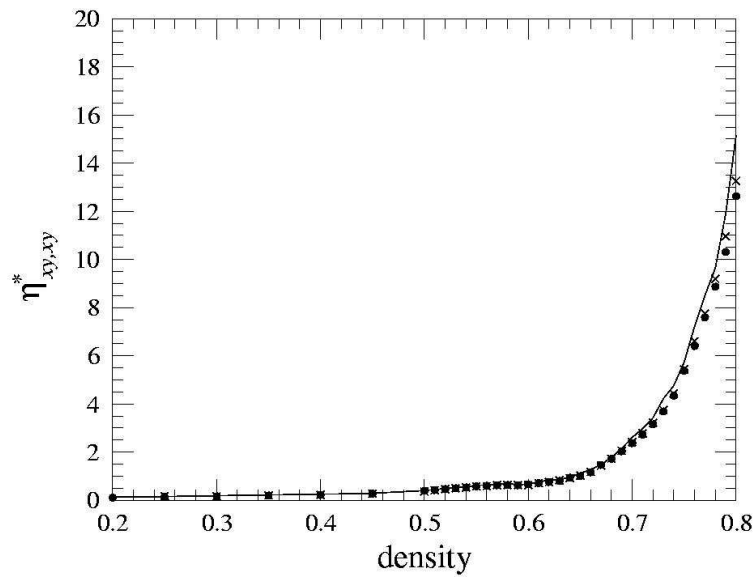,width=10cm}}
\vspace*{0.2cm}
\caption{Comparison between the three methods calculating $\eta_{xy,xy}^{*}$ in the
square geometry: the Einstein-Helfand formula (\ref{Helfand})
(continuous line), the escape-transport formula (\ref{visc-esc})
(dots), and the chaos-transport formula (\ref{visc-Hcodim}) (crosses)
with $\chi=45\sqrt{n}$ for density $n<0.66$, $\chi=100\sqrt{n}$ for $0.67<n<0.75$, and
$\chi=150\sqrt{n}$ for $0.76<n$.}
\label{visc-helf-taux-dim-carre_dens}
\end{figure}

By varying the parameter $\chi$, we have obtained the shear viscosity thanks to the
chaos-transport formula (\ref{visc-Hcodim}).  These values are plotted in Figs.
\ref{visc-helf-taux-dim-hex_dens} and
\ref{visc-helf-taux-dim-carre_dens} for the hexagonal
and square geometries, respectively.  We consider the shear viscosity
coefficient $\eta^{*}$ in the hexagonal geometry and the element
$\eta_{xy,xy^{*}}$ of the
viscosity tensor in the square geometry.  The values obtained with the
chaos-transport formula
(\ref{visc-Hcodim}) are compared with the values obtained by the
escape-transport formula
(\ref{visc-esc}) and those by the Einstein-Helfand formula (\ref{Helfand})
obtained in the
previous paper \cite{premier-article}. The agreement between the three formulas is
excellent, which confirms the theoretical results.

\vspace*{0.5cm}

\section{Conclusions}
\label{conclusions}

In the present paper, we have applied the escape-rate formalism to the
computation of
shear viscosity in the two-disk model by Bunimovich and Spohn
\cite{spohn-buni}.

The escape-rate formalism implies the appearance of a fractal repeller
associated with viscosity.
We have numerically generated the fractal repeller associated with
viscosity in this model and we
have provided evidence for its fractal character. Using the chaos-transport
formula of the
escape-rate formalism, we have been able to evaluate the shear viscosity
from the positive
maximum Lyapunov exponent and the Hausdorff codimension of the fractal
repeller of viscosity.
The values obtained by using the chaos-transport formula for shear
viscosity have been compared
with the values obtained by other methods based on the Einstein-Helfand
formula, which is
equivalent to the Green-Kubo formula as shown in our previous paper
\cite{premier-article}.  An
excellent agreement has been observed between the different methods.  This
agreement brings an
important support to the escape-rate formalism as a method to establish a
connection between the
transport properties -- here of viscosity -- and the underlying microscopic
chaotic dynamics.
The agreement therefore confirms the theoretical results of the escape-rate
formalism
\cite{gasp-nicolis,dorf-gasp}.

This confirmation opens perspectives for our understanding of the
hydrodynamic modes of shear
viscosity.  Indeed, in the case of diffusion, the fractal character of the
repeller has been shown
to imply that the nonequilibrium states are described by singular measures
defined in the
phase space of the system.  These results suggest a similar scenario in the
case of viscosity that
the nonequilibrium states of a sheared fluid should also be described by
singular measures at the
phase-space level of description.

Our results can be extended to investigate the connection between viscosity
and underlying chaos
in many-particle systems with a whole spectrum of positive Lyapunov
exponents.  In many-particle
systems, the fractal repeller is characterized by a spectrum of partial
fractal dimensions which
enter the chaos-transport formula (\ref{visc-codim2}).  In this way, we can
decompose a transport
property such as viscosity onto the spectrum of Lyapunov exponents.

The escape-rate formalism also opens perspectives to study viscosity and other
transport properties in nanoscopic systems such as nanopipes or atomic and
molecular clusters.
Indeed, the escape-rate formalism provides a way to define the transport
properties already in
nanoscopic systems containing a very small number of particles.  The
escape-rate formalism is
particularly appropriate for nanoscopic systems because the transport
coefficients can be defined
with the escape rate of the Helfand moment out of a finite range of
variation.  Since the
Helfand moment for viscosity can be seen as the center of momenta of the
particles its
variations are limited to a finite range in a system containing a finite
number of particles.
In this way, the transport properties could be studied for nanoscopic
systems of increasing size
in a similar way as the equilibrium thermodynamic properties have been
studied in nanoscopic
systems as a function of their size.

We hope to report on these issues in future publications.

\vskip 0.5 cm

{\bf Acknowledgments.}
The authors thank Professors J. R. Dorfman and G. Nicolis for support and
encouragement in this
research, and Dr. I. Claus for helpful discussions. SV thanks the FRIA for
financial support.
PG thanks the National Fund for Scientific Research (FNRS Belgium) for
financial support.


\end{document}